\documentclass[vecphys]{svmult}

\usepackage{makeidx}         
\usepackage{graphicx}        
\usepackage{multicol}        
\usepackage[bottom]{footmisc}

\usepackage{amsmath,amssymb,exscale}

\makeindex             

\def\mco{\multicolumn}
\newcommand{\be}{\begin{equation}}
\newcommand{\ee}{\end{equation}}
\newcommand{\ba}{\begin{eqnarray}}
\newcommand{\ea}{\end{eqnarray}}
\def\simlt{\mathrel{\lower2.5pt\vbox{\lineskip=0pt\baselineskip=0pt
             \hbox{$<$}\hbox{$\sim$}}}}
\def\simgt{\mathrel{\lower2.5pt\vbox{\lineskip=0pt\baselineskip=0pt
             \hbox{$>$}\hbox{$\sim$}}}}

\def\Journal#1#2#3#4{{#1} {\bf #2}, #3 (#4)}

\def\NPB{{\em Nucl. Phys.} B}
\def\PLB{{\em Phys. Lett.}  B}
\def\PRL{\em Phys. Rev. Lett.}
\def\PRD{{\em Phys. Rev.} D}

\def\JHEP{{\em JHEP} }
\def\RMP{\em Rev. Mod. Phys.}
\def\CQG{\em Class. Quant. Grav.}
\def\NJP{\em New Jour. Phys.}

\begin{document}

\title*{The Physics of Extra Dimensions}
\author{I.~Antoniadis\inst{1}}
\institute{Department of Physics, CERN - Theory Division, 1211 Geneva 23, Switzerland
\texttt{ignatios.antoniadis@cern.ch\footnote{On leave from CPHT 
(UMR CNRS 7644) Ecole Polytechnique, F-91128 Palaiseau}}}
%
%
\maketitle

Lowering the string scale in the TeV region provides a theoretical framework for solving the mass hierarchy problem and unifying all interactions. The apparent weakness of gravity can then be accounted by the existence of large internal dimensions, in the submillimeter region, and transverse to a braneworld where our universe must be confined. I review the main properties of this scenario and its implications for observations at both particle colliders, and in non-accelerator gravity experiments. Such effects are for instance the production of Kaluza-Klein resonances, graviton emission in the bulk of extra dimensions, and a radical change of gravitational forces in the submillimeter range. I also discuss the warped case and localization of gravity in the presence of infinite size extra dimensions.

\section{Introduction}
\vskip 0.2cm

During the last few decades, physics beyond the Standard Model (SM) was guided
from the problem of mass hierarchy. This can be formulated as the question of
why gravity appears to us so weak compared to the other three known fundamental 
interactions corresponding to the electromagnetic, weak and strong nuclear
forces. Indeed, gravitational interactions are suppressed by a very high energy
scale, the Planck mass $M_P\sim 10^{19}$ GeV, associated to a length
$l_P\sim 10^{-35}$ m, where they are expected to become
important. In a quantum theory, the hierarchy implies a severe fine tuning of the 
fundamental parameters in more than 30 decimal places in order to keep the 
masses of elementary particles at their observed values. The reason is that 
quantum radiative corrections to all masses generated by the 
Higgs vacuum expectation value (VEV) are proportional to the ultraviolet cutoff 
which in the presence of gravity is fixed by the Planck mass. As a result, all masses 
are ``attracted" to become about $10^{16}$ times heavier than their observed values.

Besides compositeness, there are three main theories that have been proposed
and studied extensively during the last years, corresponding to different 
approaches of dealing with the mass hierarchy problem. (1) Low energy
supersymmetry with all superparticle masses in the TeV region. Indeed, in the
limit of exact supersymmetry, quadratically divergent corrections to the Higgs
self-energy are exactly cancelled, while in the softly broken case, they are
cutoff by the supersymmetry breaking mass splittings. (2) TeV scale strings,
in which quadratic divergences are cutoff by the string scale and low
energy supersymmetry is not needed. (3) Split supersymmetry, where
scalar masses are heavy while fermions (gauginos and higgsinos) are light.
Thus, gauge coupling unification and dark matter candidate are preserved
but the mass hierarchy should be stabilized by a different way and the low energy
world appears to be fine-tuned. All these ideas
are experimentally testable at high-energy particle colliders and in
particular at LHC. Below, I discuss their implementation in string theory.

The appropriate and most convenient framework for low energy supersymmetry
and grand unification is the perturbative heterotic string. Indeed, in this theory,
gravity and gauge interactions have the same origin, as massless modes of the
closed heterotic string, and they are unified at the string scale $M_s$. As a result, 
the Planck mass $M_P$ is predicted to be proportional to $M_s$:
\be
M_P=M_s/g\, ,
\label{het}
\ee
where $g$ is the gauge coupling. In the simplest constructions all gauge 
couplings are the same at the string scale, given by the four-dimensional (4d)
string coupling, and thus no grand unified group is needed for unification.
In our conventions $\alpha_{\rm GUT}=g^2\simeq 0.04$, leading to a 
discrepancy between the string and grand unification scale 
$M_{\rm GUT}$ 
by almost two orders of magnitude. Explaining this gap introduces in general
new parameters or a new scale, and the predictive power is essentially
lost. This is the main defect of this framework, which remains though an open
and interesting possibility.

\begin{figure}[htb]
\begin{center}
\includegraphics[width=8cm]{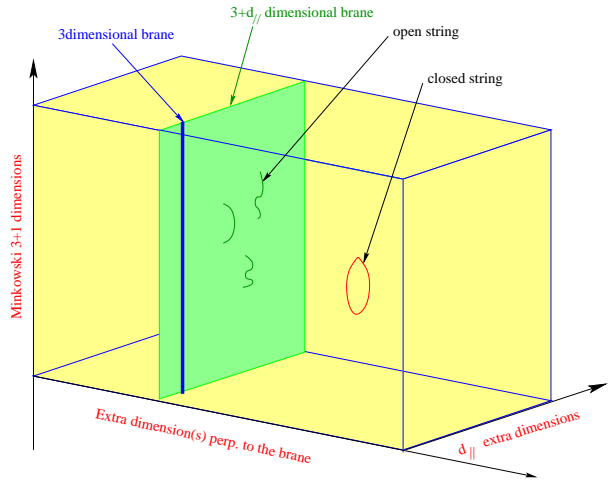}
\end{center}
\caption{\small In the type I string framework, our Universe contains,
besides the three known spatial dimensions (denoted by a single blue
line), some extra dimensions ($d_\parallel=p-3$) parallel to our world
$p$-brane (green plane) where endpoints of open strings are confined,
as well as some transverse dimensions (yellow space) where
only gravity described by closed strings can propagate.
\label{model}}
\end{figure}

The other two ideas have both as natural framework of realization type I
string theory with D-branes. Unlike in the heterotic string, gauge and gravitational
interactions have now different origin. The latter are described again by closed
strings, while the former emerge as excitations of open strings with
endpoints confined on D-branes~\cite{Angelantonj:2002ct}. 
This leads to a braneworld description of our universe, which
should be localized 
on a hypersurface, i.e. a membrane extended in $p$ spatial dimensions, 
called $p$-brane (see Fig.~\ref{model}). Closed strings propagate in all nine 
dimensions of string theory: in those extended along the $p$-brane, called parallel, 
as well as in the transverse ones. On the contrary, open strings are attached 
on the $p$-brane.
Obviously, our $p$-brane world must have at least the three known
dimensions of space. But it may contain more: the extra $d_\parallel=p-3$
parallel dimensions must have a finite size, in order to be unobservable at 
present energies, and can be as large as TeV$^{-1}\sim 10^{-18}$
m~\cite{ia}. On the other hand, transverse dimensions interact with us only 
gravitationally and experimental bounds are much weaker: their size should 
be less than about 0.1 mm~\cite{Hoyle:2004cw}. In the following, I review
the main properties and experimental signatures of low string 
scale models~\cite{aadd,Antoniadis:2002da}.

\section{Framework}
\vskip 0.2cm

In type I theory, the different origin of gauge and  gravitational interactions
implies that the relation between the Planck and string scales is not linear as
(\ref{het}) of the heterotic string. The requirement that string theory
should be weakly coupled, constrain the size of all parallel dimensions to be of
order of the string length, while transverse dimensions remain unrestricted.
Assuming an isotropic transverse space of $n=9-p$ compact 
dimensions of common radius $R_\perp$, one finds:
\begin{equation}
M_P^2=\frac{1}{g^4}M_s^{2+n}R_\perp^n\ ,\qquad
g_s \simeq g^2\, .
\label{treei}
\end{equation}
where $g_s$ is the string coupling. It follows that the type I
string scale can be chosen hierarchically smaller than the Planck 
mass~\cite{Lykken:1996fj,aadd} at
the expense of introducing extra large transverse dimensions
felt only by gravity, while keeping the string coupling small~\cite{aadd}. 
The weakness of 4d gravity compared to gauge interactions
(ratio $M_W/M_P$) is then
attributed to the largeness of the transverse space $R_\perp$
compared to the string length $l_s=M_s^{-1}$. 

\begin{figure}[htb]
\begin{center}
\includegraphics[width=7.5cm]{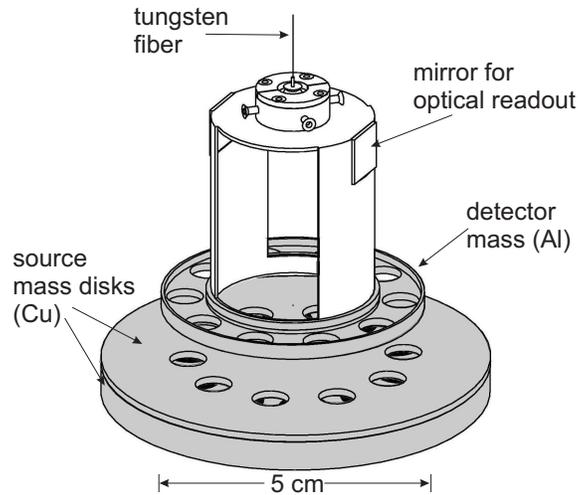}
\end{center}
\caption{\small Torsion pendulum that tested 
Newton's law at 130 nm. 
Several sources of background noise were eliminated
using appropriate devices.
\label{newton}}
\end{figure}

An important property of these models is that gravity becomes effectively
$(4+n)$-dimensional  with a strength comparable to those of gauge
interactions at the string scale. The first relation of
Eq.~(\ref{treei}) can be understood as a consequence of the
$(4+n)$-dimensional Gauss law for gravity, with 
\be
M_*^{(4+n)}=M_s^{2+n}/g^4 
\label{GN}
\ee
the effective scale of gravity in $4+n$ dimensions.
Taking $M_s\simeq 1$ TeV,
one finds a size for the extra dimensions $R_\perp$ varying from
$10^8$ km, .1 mm, down  to a Fermi for $n=1,2$,
or 6 large dimensions, respectively. This shows that while $n=1$ is 
excluded, $n\geq 2$ is allowed by present experimental bounds
on gravitational forces~\cite{Hoyle:2004cw,newtonslaw}. Thus, in these models, gravity
appears to us very weak at macroscopic scales because its intensity is spread in the
``hidden" extra dimensions. At distances shorter than $R_\perp$, 
it should deviate from
Newton's law, which may be possible to explore in laboratory experiments 
(see Fig.~\ref{newton}).

The main experimental implications of TeV scale strings in particle
accelerators are of three types, in correspondence with the three different
sectors that are generally present: (i) new compactified parallel dimensions,
(ii) new extra large transverse dimensions and low scale quantum gravity, and
(iii) genuine string and quantum gravity effects. On the other hand, there
exist interesting implications in non accelerator table-top experiments
due to the exchange of gravitons or other possible states living in the bulk.

\section{Experimental implications in accelerators}
\vskip 0.2cm
\subsection{World-brane extra dimensions}
\vskip 0.2cm

In this case $RM_s\simgt 1$, and the associated compactification scale 
$R^{-1}_\parallel$ would be the first scale of new physics that should be 
found increasing the beam energy~\cite{ia,iab}.
There are several reasons for the existence of such dimensions.
It is a logical possibility, since out of the six extra dimensions of string theory
only two are needed for lowering the string scale, and thus the effective 
$p$-brane of our world has in general $d_\parallel\equiv p-3\le 4$.
Moreover, they can be used to address several physical problems in 
braneworld models, such as obtaining
different SM gauge couplings, explaining fermion mass hierarchies
due to different localization points of quarks and leptons in the extra 
dimensions, providing calculable mechanisms of supersymmetry breaking, etc.

The main consequence is the existence of Kaluza-Klein (KK) excitations
for all SM particles that propagate along the extra parallel dimensions.
Their masses are given by:
\be
M_m^2=M_0^2+{m^2\over R_\parallel^2}\quad ;\quad
m=0,\pm 1,\pm 2,\dots
\label{KKmass}
\ee
where we used $d_\parallel=1$, and $M_0$ is the higher dimensional mass.
The zero-mode $m=0$ is identified with the 4d state, while
the higher modes have the same quantum numbers with the lowest one, 
except for their mass given in (\ref{KKmass}).
There are two types of experimental signatures of such 
dimensions~\cite{iab,abq,AAB}:
(i) virtual exchange of KK excitations, leading to
deviations in cross-sections compared to the SM prediction, that can be used
to extract bounds on the compactification scale;
(ii) direct production of KK modes.

On general grounds, there can be two different kinds of models with
qualitatively different signatures depending on the localization properties 
of matter fermion fields. If the latter are localized in 3d
brane intersections, they do not have excitations and
KK momentum is not conserved because of the breaking of translation
invariance in the extra dimension(s). KK modes of gauge 
bosons are then singly produced giving rise to generally strong bounds on
the compactification scale and new resonances that can be observed in 
experiments. Otherwise, they can be produced only in pairs 
due to the KK momentum conservation,
making the bounds weaker but the resonances difficult to observe.

When the internal momentum is conserved, the interaction vertex involving 
KK modes has the same 4d tree-level gauge coupling.
On the other hand, their couplings to localized matter have an exponential
form factor suppressing the interactions of heavy modes.
This form factor can be viewed as the fact that the branes intersection has a
finite thickness. 
For instance, the coupling of the KK excitations of gauge fields
$A^\mu (x, y)=\sum_m
A^{\mu }_m \exp{i\frac {m y}{R_\parallel}}$ to the charge
density $j_\mu (x)$ of massless  localized fermions is described
by the effective action~\cite{ABL}: 
\be
\int d^4x \, \sum_ m e^{-\ln {16}\frac{m^2l_s^2}{2 R_\parallel^2}} \, 
j_\mu (x) \, A^{\mu }_m(x)\, .
\label{momwidth}
\ee
After Fourier transform in position space, it becomes:
\be 
\int d^4x\, dy\, 
\frac{1}{(2 \pi \ln 16)^2} e^{-\frac{y^2M_s^2}{2\ln 16}}\, 
j_\mu (x) \, A^\mu (x, y)\, ,
\label{brwidth}
\ee
from which we see that localized fermions form 
a Gaussian distribution of charge 
with a width $\sigma=\sqrt{\ln 16}\, l_s \sim 1.66 \, l_s$.

To simplify the analysis, let us consider first the case $d_\parallel=1$ 
where some of the gauge fields arise from an effective 4-brane, while 
fermions are localized states on brane intersections. 
Since the corresponding gauge couplings are reduced 
by  the size of the large dimension $R_\parallel M_s$ compared to 
the others, one can account for the ratio of the weak to strong interactions
strengths if the $SU(2)$ brane extends along the extra dimension, while
$SU(3)$ does not. 
As a result, there are 3 distinct cases to study~\cite{AAB},
denoted by $(t,l,l)$, $(t,l,t)$ and $(t,t,l)$, where the three
positions in the brackets correspond to the three SM gauge group factors
$SU(3)\times SU(2)\times U(1)$ and those with $l$ (longitudinal) feel
the extra dimension, while those with $t$ (transverse) do not.

In the $(t,l,l)$ case, there are KK excitations of $SU(2)\times U(1)$
gauge bosons: $W_{\pm}^{(m)}$, $\gamma^{(m)}$ and $Z^{(m)}$.
Performing a $\chi^2$ fit of the electroweak observables, 
one finds that if the Higgs is a bulk state $(l)$,
$R_\parallel^{-1} \simgt 3.5$ TeV~\cite{Delgado}. 
This implies that LHC can produce at most the first KK mode. 
Different choices for localization of matter and  Higgs fields lead to 
bounds, lying in the range $1-5$ TeV~\cite{Delgado}.

\begin{figure}[htb]
\begin{center}
\includegraphics[width=8cm]{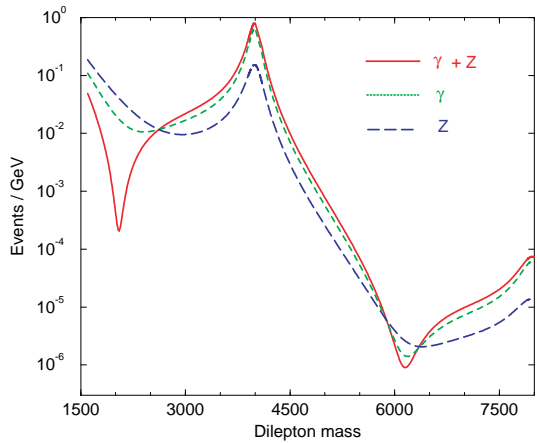}
\end{center}
\caption{\small Production of the first KK modes of the
photon and of the $Z$ boson at LHC, decaying to electron-positron pairs. 
The number of expected events is plotted
as a function of the energy of the pair in GeV. 
From highest to lowest: excitation of
$\gamma+Z$, $\gamma$ and $Z$.
\label{KK}}
\end{figure}

In addition to virtual effects, KK excitations can be produced
on-shell at LHC as new resonances~\cite{abq} (see Fig.~\ref{KK}). 
There are two different channels,
neutral Drell--Yan processes $pp \rightarrow l^+l^-X$ and the charged 
channel $l^{\pm} \nu$, corresponding to the production of the KK modes
$\gamma^{(1)},Z^{(1)}$ and $W_\pm^{(1)}$, respectively.
The discovery limits are about 6 TeV, while the exclusion bounds
15 TeV. An interesting observation in the
case of $\gamma^{(1)} + Z^{(1)}$ is that interferences can lead
to a ``dip'' just before the resonance. 
There are some ways to distinguish the corresponding signals from other
possible origin of new physics, such as models with new gauge bosons. 
In fact, in the $(t,l,l)$ and $(t,l,t)$ cases, one expects two resonances located
practically at the same mass value. This property is not shared by most of
other new gauge boson models. Moreover, the heights and widths of the
resonances are directly related to those of SM gauge bosons in
the corresponding channels.

In the $(t,l,t)$ case, only the $SU(2)$ factor feels the
extra dimension  and the limits set by the KK states of $W^{\pm}$
remain the same. On the other hand,
in the $(t,t,l)$ case where only $U(1)_Y$ feels 
the extra dimension, the limits are
weaker and the exclusion bound is around 8 TeV.
In addition to these simple possibilities, brane constructions lead often 
to cases where part of $U(1)_Y$ is $t$ and part is $l$. If
$SU(2)$ is $l$ the limits come again from $W^{\pm}$, while if it is $t$
then it will be difficult to distinguish this case from a generic extra $U(1)'$. 
A good statistics would be needed to see the deviation in the tail of the
resonance as being due to effects additional to those of a generic $U(1)'$
resonance. Finally, in the case of two or more parallel dimensions, 
the sum in the exchange of the KK modes diverges in the limit 
$R_\parallel M_s>>1$ and needs to be regularized using the form 
factor~(\ref{momwidth}). Cross-sections become bigger yielding stronger 
bounds, while resonances are closer implying that more of them 
could be reached by LHC.

On the other hand, if all SM particles propagate in the extra dimension
(called universal)\footnote{Although interesting, this scenario seems difficult
to be realized, since 4d chirality requires non-trivial action of orbifold twists 
with localized chiral states at the fixed points.}, 
KK modes can only be produced in pairs and the
lower bound on the compactification scale becomes weaker, of order of 
300-500 GeV. Moreover, no resonances can be observed at LHC, so that
this scenario appears very similar to low energy supersymmetry. In fact,
KK parity can even play the role of R-parity, implying that the lightest KK
mode is stable and can be a dark matter candidate in analogy 
to the LSP~\cite{Servant:2002aq}.

\subsection{Extra large transverse dimensions}
\vskip 0.2cm

The main experimental signal is gravitational radiation in the bulk from
any physical process on the world-brane. In fact, the very existence of
branes breaks translation invariance in the transverse dimensions and 
gravitons can be emitted from the brane into the bulk.
During a collision of center of mass energy $\sqrt{s}$, there are 
$\sim (\sqrt{s}R_{\perp})^n$ KK excitations of gravitons with tiny masses,
that can be emitted. Each of these states 
looks from the 4d point of view as a massive, quasi-stable, 
extremely weakly coupled ($s/M^2_P$ suppressed) particle that escapes
from the detector. The total effect is a missing-energy cross-section
roughly of order: 
\be
\frac {(\sqrt{s}R_{\perp })^n} {M^2_P} \sim \frac{1}{s} 
{\left(\frac{\sqrt{s}}{M_s}\right)^{n+2}}\, .
\label{miss1}
\ee
Explicit computation of these effects leads to the bounds given in 
Table~\ref{tab:exp3}.
\begin{table}[tb]
\centering
\caption{\label{tab:exp3} Limits on $R_\perp$ in mm.} 
\vskip 0.2 cm
\small
\begin{tabular}{  | c | c | c | c |} 
\hline
  & & & \\   
{\tiny Experiment} & $n=2$ & $n=4$ & $n=6$ \\ 
\hline\hline
\mco{4}{|c|}{Collider bounds}   \\ \hline

 LEP 2   & $5\times 10^{-1}$ & $2\times 10^{-8}$  & 
                              $7 \times 10^{-11}$ \\ \hline
  Tevatron  &   $5 \times 10^{-1}$  & $10^{-8}$ 
              & $4 \times 10^{-11}$ \\ \hline 
  LHC &  $4 \times 10^{-3}$   & $6\times 10^{-10}$  & 
                              $3 \times 10^{-12}$  \\ \hline
  NLC & $ 10^{-2}$  & $10^{-9}$  & 
                              $6 \times 10^{-12}$  \\ \hline\hline
\mco{4}{|c|}{Present non-collider bounds}   \\ \hline
  
{\tiny SN1987A}   &  $3 \times 10^{-4}$   & 
           $ 10^{-8}$ 
                 & $6 \times 10^{-10} $ \\ \hline
{\tiny COMPTEL} &  $5 \times 10^{-5}$   & - & 
                              - \\ \hline
\end{tabular}
\end{table}
However, larger radii are allowed if one relaxes the assumption of isotropy, 
by taking for instance two large dimensions with different radii.

\begin{figure}[htb]
\begin{center}
\includegraphics[width=8cm]{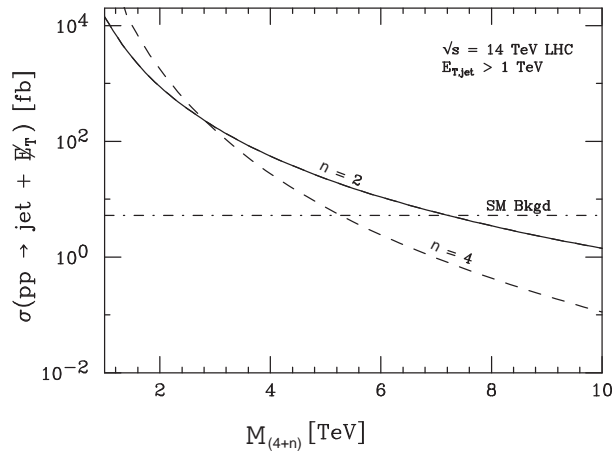}
\end{center}
\caption{\small Missing energy due to graviton emission at LHC, as a
function of the higher-dimensional gravity scale $M_*$,
produced together with a hadronic jet. The expected cross-section is shown 
for $n=2$ and $n=4$ extra dimensions, together with the SM background.
\label{graviton}}
\end{figure}

{}Fig.~\ref{graviton} shows the
cross-section for graviton emission in the bulk, corresponding to the
process $pp\to jet + graviton$ at LHC, together with the SM
background~\cite{missing}. 
{}For a given value of $M_s$, the cross-section for graviton emission
decreases with the number of large transverse dimensions, in contrast
to the case of parallel dimensions. The reason is that gravity becomes
weaker if there are more dimensions because there is more space for
the gravitational field to escape.
There is a particular energy and angular distribution  of the
produced gravitons that arise from the distribution in mass  of KK
states of spin-2. This can be contrasted to other sources of missing 
energy and might be a smoking gun for the
extra dimensional nature of such a signal.

In Table~\ref{tab:exp3}, there are also included astrophysical and cosmological
bounds. Astrophysical bounds~\cite{add2,supernovae} arise from the
requirement that the radiation of gravitons should not carry on too much
of the gravitational binding energy released during core collapse of
supernovae. In fact, the measurements of Kamiokande and IMB for SN1987A 
suggest that the main channel is neutrino fluxes.
The best cosmological bound~\cite{COMPTEL} is obtained from requiring that
decay of bulk gravitons to photons do not generate a spike in the energy
spectrum of the photon background measured by the COMPTEL instrument. Bulk 
gravitons  are  expected  to be produced just before
nucleosynthesis due to thermal radiation from the brane. The limits assume
that the temperature was at most 1 MeV as nucleosynthesis begins, and become
stronger if temperature is increased.

\subsection{String effects}
\vskip 0.2cm

At low energies, the interaction of light (string) states is described by an
effective field theory. Their exchange generates in particular four-fermion operators
that can be used to extract independent bounds on the string scale. 
In analogy with the bounds on longitudinal extra dimensions, there are
two cases depending on the localization properties of matter fermions.
If they come from open strings with both ends on the same stack of branes,
exchange of massive open string modes gives rise to dimension eight effective
operators, involving four fermions and two space-time derivatives~\cite{Peskin,ABL}. 
The corresponding bounds on the string scale are then around 500 GeV.
On the other hand, if matter fermions are localized on non-trivial brane intersections,
one obtains dimension six four-fermion operators and the bounds become
stronger: $M_s\simgt 2-3$ TeV~\cite{ABL,Antoniadis:2002da}.
At energies higher than the string scale, new spectacular phenomena are
expected to occur, related to string physics and quantum gravity effects,
such as possible micro-black hole production~\cite{bh}. 
Particle accelerators would then become the 
best tools for studying quantum gravity and string theory.

\section{Supersymmetry in the bulk and short range forces}
\vskip 0.2cm
\subsection{Sub-millimeter forces}
\vskip 0.2cm
 
Besides the spectacular predictions in
accelerators, there are also modifications of gravitation in
the sub-millimeter range, which can be tested in ``table-top"
experiments that measure gravity at short distances. There are three
categories of such predictions:\hfil\\  
(i) Deviations from the Newton's law $1/r^2$ behavior to $1/r^{2+n}$, 
which can be observable for $n=2$ 
large transverse dimensions of sub-millimeter size. 
This case is particularly attractive
on theoretical grounds because of the logarithmic sensitivity of SM
couplings on the size of transverse space~\cite{ab}, that allows
to determine the hierarchy~\cite{abml}.\hfil\\ 
(ii) New scalar
forces in the sub-millimeter range, related to the mechanism of
supersymmetry breaking, and mediated by light scalar fields
$\varphi$ with masses~\cite{iadd,aadd}:
\be
m_{\varphi}\simeq{m_{susy}^2\over M_P}\simeq 
10^{-4}-10^{-6}\ {\rm eV} \, ,
\label{msusy}
\ee
for a supersymmetry breaking scale $m_{susy}\simeq 1-10$ TeV. They
correspond to Compton wavelengths of 1 mm to 10 $\mu$m.
$m_{susy}$ can be either $1/R_\parallel$ if supersymmetry is broken by
compactification~\cite{iadd}, or the string scale if it is broken
``maximally" on our world-brane~\cite{aadd}. 
A universal attractive scalar force is mediated by the radion modulus
$\varphi\equiv M_P\ln R$,
with $R$ the radius of the longitudinal or transverse dimension(s).
In the former case, the result
(\ref{msusy}) follows from the behavior of the vacuum energy density
$\Lambda \sim 1/R^4_\parallel$ for large $R_\parallel$ (up to logarithmic
corrections). In the latter, supersymmetry is broken primarily on
the brane, and thus its transmission to the bulk is gravitationally
suppressed, leading to (\ref{msusy}). For $n=2$,
there may be an enhancement factor of the radion
mass by $\ln R_\perp M_s\simeq 30$ decreasing its wavelength by
an order of magnitude~\cite{abml}.

\begin{figure}[htb]
\begin{center}
\includegraphics[width=8cm]{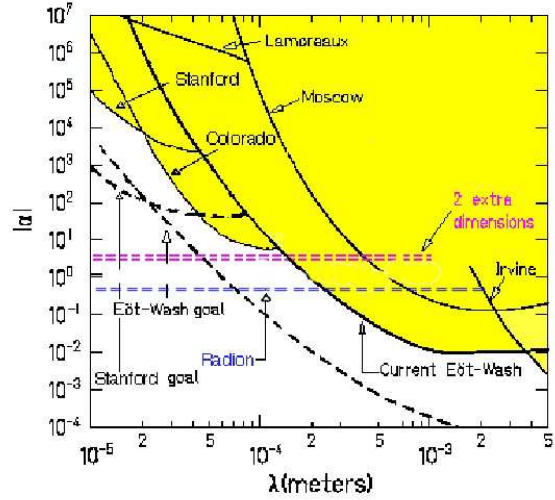}
\end{center}
\caption{\small Present limits on non-Newtonian forces at short distances
(yellow regions), as a function of their range $\lambda$ 
and their strength relative to gravity $\alpha$.
The limits are compared to new forces mediated by the graviton in the
case of two large extra dimensions, and by the radion.
\label{forces}}
\end{figure}

The coupling of the radius modulus 
to matter relative to
gravity can be easily computed and is given by:
\be\small
\sqrt{\alpha_\varphi} = {1\over M}{\partial M\over\partial\varphi}\ 
;\ \ \alpha_\varphi=\left\{ \begin{array}{l}
{\partial\ln\Lambda_{\rm QCD}\over\partial\ln R}\simeq {1\over 3}\ \ 
{\rm for}\ R_\parallel\\ \\
{2n\over n+2}=1 - 1.5\ {\rm for}\ R_\perp
\end{array}\right.
\label{dcoupling}
\ee
where $M$ denotes a generic physical mass. In the longitudinal case,
the coupling arises dominantly through the radius
dependence of the QCD gauge coupling~\cite{iadd}, while in the
case of transverse dimension, it can be deduced from the rescaling of the
metric which changes the string to the Einstein frame and depends slightly on the
bulk dimensionality ($\alpha=1-1.5$ for $n=2-6$)~\cite{abml}. 
Such a force can be tested in
microgravity experiments and should be contrasted with the change of
Newton's law due the presence of extra dimensions that is observable only
for $n=2$~\cite{Hoyle:2004cw,newtonslaw}.
The resulting bounds from an analysis of the radion effects
are~\cite{Hoyle:2004cw}:
\be
M_*\simgt 3-4.5\, {\rm TeV}\ \ {\rm for}\ \ n=2-6\, .
\label{boundsradion}
\ee
In principle there can be other light moduli which couple with even
larger strengths. For example the dilaton, whose VEV determines
the string coupling, if it does not acquire
large mass from some dynamical supersymmetric mechanism, can lead to a
force of strength 2000 times bigger than gravity~\cite{tvdil}.\hfil\\
(iii) Non universal repulsive forces much stronger than gravity, mediated
by possible abelian gauge fields in the bulk~\cite{add2,akr}. Such
fields acquire tiny masses of the order of $M_s^2/M_P$, as in
(\ref{msusy}), due to brane localized anomalies~\cite{akr}. Although
their gauge coupling is infinitesimally small, $g_A\sim
M_s/M_P\simeq 10^{-16}$, it is still bigger that the gravitational
coupling $E/M_P$ for typical energies $E\sim 1$ GeV,
and the strength of the new force would be $10^6-10^8$ stronger than
gravity. This is an interesting region which will be soon
explored in micro-gravity experiments (see Fig.~\ref{forces}). Note
that in this case supernova constraints impose that there should be
at least four large extra dimensions in the
bulk~\cite{add2}.

\begin{figure}[htb]
\begin{center}
\includegraphics[width=9cm]{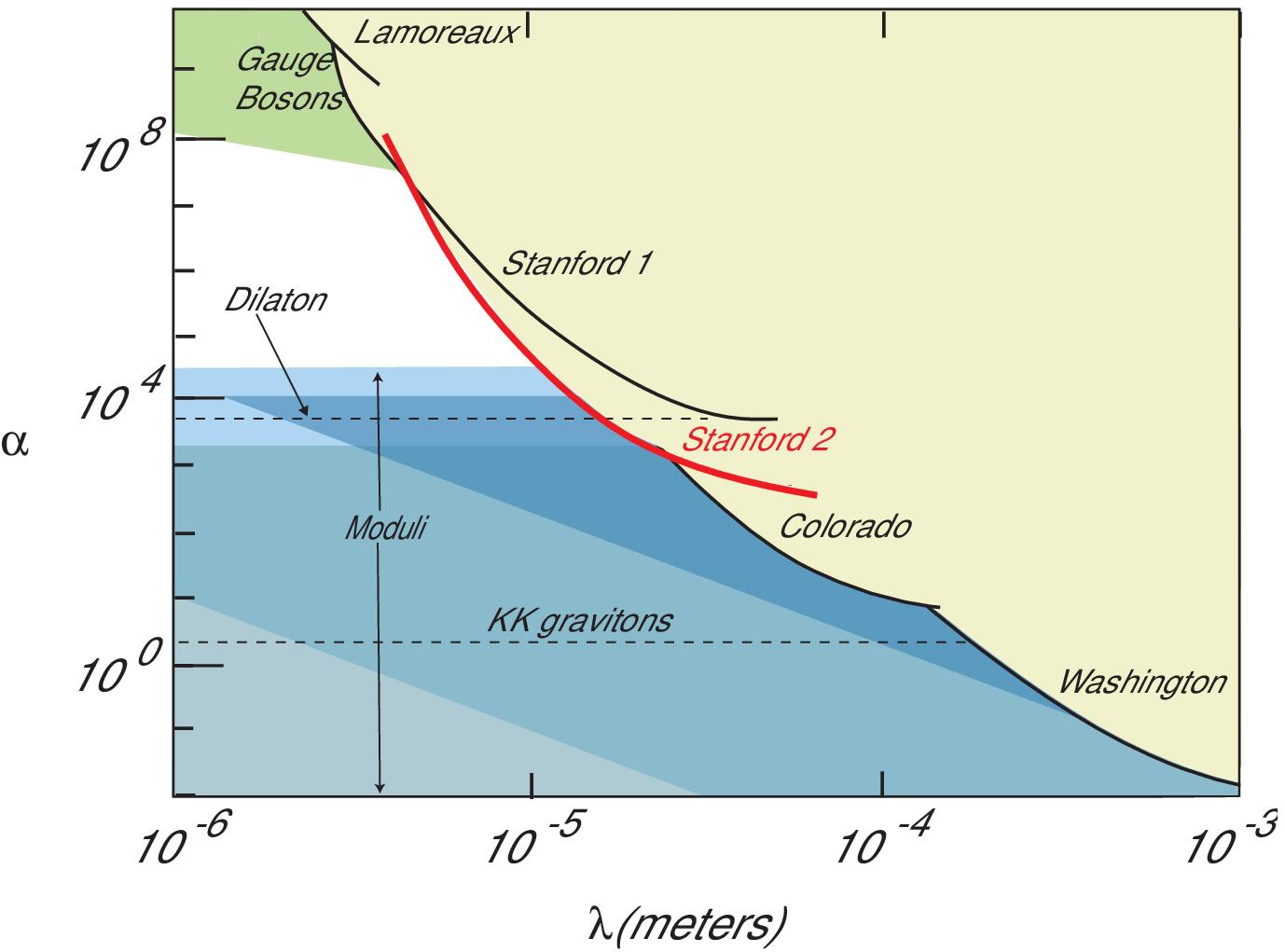}
\end{center}
\caption{\small Bounds on non-Newtonian forces in the range 6-20 $\mu$m
(see S.~J.~Smullin et al. in Ref.~\cite{newtonslaw}).
\label{forcessub1}}
\end{figure}

\begin{figure}[htb]
\begin{center}
\includegraphics[width=8cm]{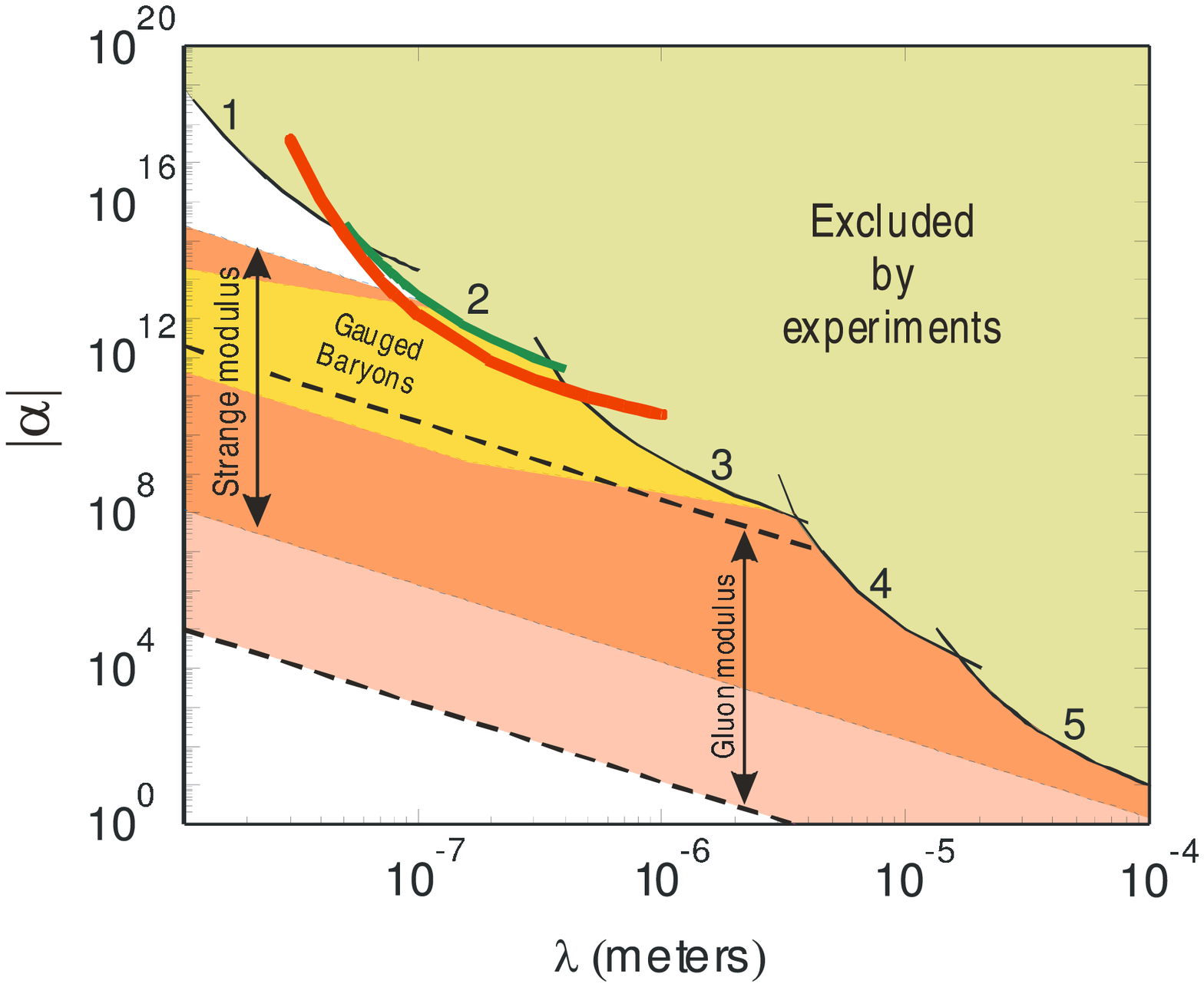}
\end{center}
\caption{\small Bounds on non-Newtonian forces in the range around 200 nm
(see R.~S.~Decca et al. in Ref.~\cite{newtonslaw}). Curves 4 and 5 correspond
to Stanford and Colorado experiments, respectively, of Fig.~\ref{forcessub1}
(see also J~C.~Long and J.~C.~Price of Ref.~\cite{newtonslaw}).
\label{forcessub2}}
\end{figure}

In Fig.~\ref{forces} we depict the actual information from previous,
present and upcoming experiments~\cite{newtonslaw,abml}. 
The solid lines indicate the present limits from the experiments indicated.
The excluded regions lie above these solid lines. Measuring gravitational
strength forces at short distances is challenging. 
The dashed thick lines give the expected
sensitivity of the various experiments, which will improve
the actual limits by roughly two orders of magnitude, while the
horizontal dashed lines correspond to the theoretical predictions for the
graviton in the case $n=2$ and for the radion in
the transverse case. These limits are compared to those
obtained from particle accelerator experiments in Table~\ref{tab:exp3}.
Finally, in Figs.~\ref{forcessub1} and \ref{forcessub2}, we display recent 
improved bounds for new forces at very short distances by focusing on the 
right hand side of Fig.~\ref{forces}, near the origin~\cite{newtonslaw}.

\subsection{Brane non-linear supersymmetry}
\vskip 0.2cm

When the closed string sector is supersymmetric, 
supersymmetry on a generic brane configuration is non-linearly realized 
even if the spectrum is not supersymmetric and brane fields have no
superpartners. The reason is that the gravitino 
must couple to a conserved current locally, implying the existence of a 
goldstino on the brane world-volume. The goldstino is exactly massless
in the infinite (transverse) volume limit and is 
expected to acquire a small mass suppressed by the volume, 
of order (\ref{msusy}). In the standard realization, its coupling to 
matter is given via the energy momentum tensor~\cite{volaku}, 
while in general there
are more terms invariant under non-linear supersymmetry that have been
classified, up to dimension eight~\cite{bfzfer,Antoniadis:2004uk}. 

An explicit computation was performed
for a generic intersection of two brane stacks, leading to three
irreducible couplings, besides the standard 
one~\cite{Antoniadis:2004uk}: two of
dimension six involving the goldstino, a matter fermion and a scalar or
gauge field, and one four-fermion operator of dimension eight. 
Their strength is set by the goldstino decay constant $\kappa$, up to 
model-independent numerical coefficients which are 
independent of the brane angles. 
Obviously, at low energies the dominant operators are
those of dimension six. In the minimal case of (non-supersymmetric) SM,
only one of these two operators may exist, that couples the
goldstino $\chi$ with the Higgs $H$ and a lepton doublet $L$:
\be
{\cal L}_\chi^{int}=2\kappa (D_\mu H)(LD^\mu\chi)+h.c.\, ,
\label{Lchi}
\ee
where the goldstino decay constant is given by the total brane tension
\begin{equation}
{1\over 2 \ \kappa^2} = N_1 \, T_1 + N_2 \, T_2 \, ;\quad
T_i ={M_s^4 \over 4 \pi^2 g_i^2} \, ,
\label{decayconst}
\end{equation}
with $N_i$ the number of branes in each stack.
It is important to notice that the effective interaction
(\ref{Lchi}) conserves the total lepton number
$L$, as long as we assign to the goldstino a total
lepton number $L(\chi)=-1$~\cite{Antoniadis:2004se}. 
To simplify the analysis, we will consider
the simplest case where (\ref{Lchi}) exists only for the
first generation and $L$ is the electron doublet~\cite{Antoniadis:2004se}.

\begin{figure}[htb]
\begin{center}
\includegraphics[width=7.5cm]{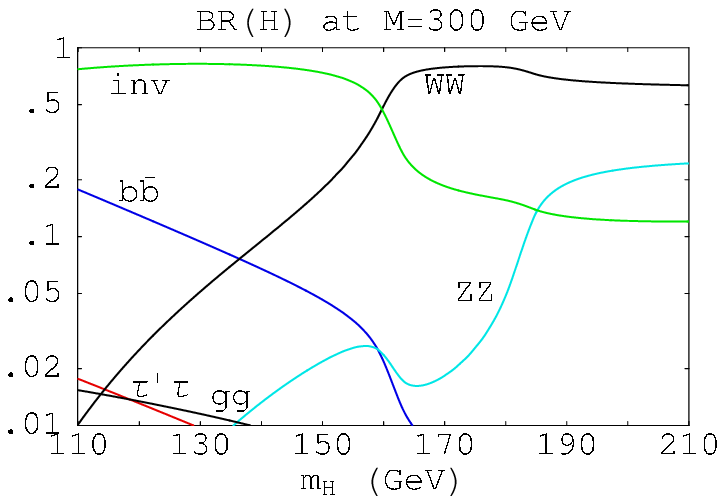}
\hskip 0.2cm
\includegraphics[width=7.5cm]{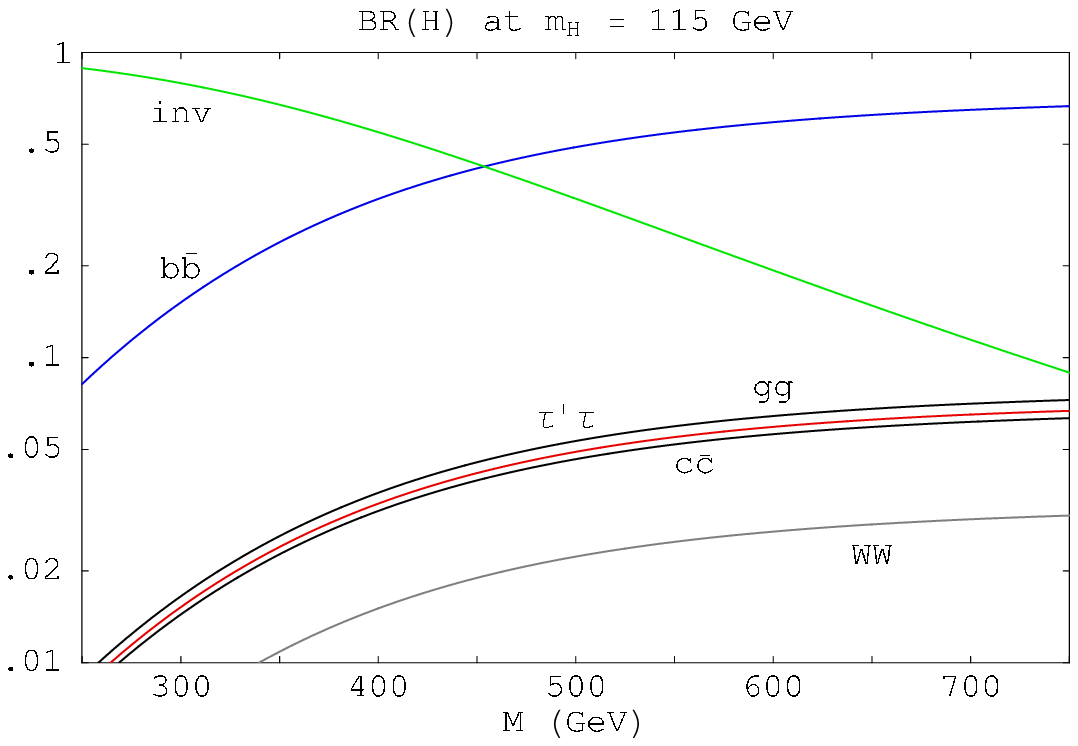}
\end{center}
\caption{\small Higgs branching rations, as functions either
of the Higgs mass $m_H$ for a fixed value of the string scale 
$M_s\simeq 2M=600$ GeV, or 
of $M\simeq M_s/2$  for $m_H=115$ GeV.
\label{Hinvisible}}
\end{figure}

The effective interaction (\ref{Lchi}) gives rise mainly to the decays
$W^\pm\to e^\pm\chi$ and $Z,H\to\nu\chi$. It turns out that the invisible
$Z$ width gives the strongest limit on $\kappa$ which can
be translated to a bound on the string scale $M_s\simgt 500$ GeV,
comparable to other collider bounds. This allows for the striking
possibility of a Higgs boson decaying dominantly, or at least with a
sizable branching ratio, via such an invisible mode, for a wide
range of the parameter space $(M_s, m_H)$, as seen in 
Fig.~\ref{Hinvisible}. 

\section{Electroweak symmetry breaking}
\vskip 0.2cm

Non-supersymmetric  TeV strings offer also a framework to realize
gauge symmetry breaking radiatively. Indeed,
from the effective field theory point of
view, one expects quadratically divergent one-loop 
contributions to the masses of scalar fields.
The divergences are cut off by $M_s$ and if the corrections are negative,
they can induce electroweak symmetry breaking and 
explain the mild hierarchy between the weak and a string scale at a few TeV,
in terms of a loop factor~\cite{abqhiggs}.
More precisely, in the minimal case of one Higgs
doublet $H$, the scalar potential is:
\be
V=\lambda (H^\dagger H)^2 + \mu^2 (H^\dagger H)\, ,
\label{potencialh}
\ee
where $\lambda$ arises at tree-level. Moreover,
in any model where the Higgs field comes from an open string 
with both ends fixed on the same brane stack, it 
is given by an appropriate truncation of a supersymmetric theory. 
Within the minimal spectrum of the SM,
$\lambda=(g_2^2+g'^2)/8$, with $g_2$ and $g'$ the $SU(2)$ and $U(1)_Y$ gauge
couplings. On the other hand, $\mu^2$ is generated at one loop: 
\be
\label{mu2R}
\mu^2=-\varepsilon^2\, g^2\, M_s^2\, ,
\ee
where $\varepsilon$ is a loop factor that can be estimated from a
toy model computation and varies in the region 
$\epsilon\sim 10^{-1}-10^{-3}$.

Indeed, consider for illustration a simple case where the whole one-loop 
effective potential of a scalar field can be computed. We assume for 
instance one extra dimension compactified on a circle of radius $R>1$ 
(in string units). An interesting situation is provided by a class of models
where a non-vanishing VEV for a scalar (Higgs) field $\phi$ results in
shifting the mass of each KK excitation by a constant $a(\phi)$: 
\be
M^2_m= \left( \frac {m+a(\phi) }{R}\right)^2\, ,
\label{mastermass}
\ee
with $m$ the KK integer momentum number. 
Such mass shifts arise for instance in the presence 
of a Wilson line, $a= q \oint \frac{dy}{2 \pi} g A$, where $ A$ is the internal
component of a gauge field with gauge coupling $g$, and $q$ is the
charge of a given state under the corresponding generator. A
straightforward computation shows that the $\phi$-dependent part of
the one-loop effective potential is given by~\cite{Antoniadis:2001cv}:
\begin{equation}
V_{eff}=- Tr (-)^{F} \, \frac{R}{32\, \pi^{3/2} }\, \, 
\sum_{n}  e^{ 2 \pi i n a}\, \,   \int_0^\infty dl \, \,  l^{3/2} f_s(l)
\,  \,  e^{- \pi^2 n^2 R^2 l}
\label{strings}
\end{equation}
where $F =0,1$ for bosons and fermions, respectively. We have included
a regulating function $f_s(l)$ which contains for example the effects
of string oscillators.  To understand its role we will consider the
two limits $R>>1$ and $R<<1$. In the first case only the
$l\rightarrow 0$ region contributes to the integral. This means that
the effective potential receives sizable contributions only from the
infrared (field theory) degrees of freedom. In this limit we would
have $f_s(l)\rightarrow 1$.  For example, in the string model
considered in~\cite{abqhiggs}:
\be
f_s(l) = \left[\frac{1}{4 l}  \frac {\theta_2} { \eta^{3}}(il+{\frac{1}{ 2}})
\right]^4 \rightarrow 1 \qquad {\rm for }\qquad l\rightarrow 0, 
\ee
and the field theory result is finite and can be explicitly computed.
As a result of the Taylor expansion around $a=0$, we are able to
extract the one-loop contribution to the coefficient of the
term of the potential quadratic in the Higgs field. It is given by a
loop factor times the compactification scale~\cite{Antoniadis:2001cv}. 
One thus obtains $\mu^2 \sim g^2/ R^2$ up to a proportionality
constant which is calculable in the effective field theory.
On the other hand, if we consider $R \rightarrow 0$, which by
$T$-duality corresponds to taking the extra dimension as transverse
and very large, the one-loop effective potential receives
contributions from the whole tower of string oscillators as appearing
in $f_s(l)$, leading to squared masses given by a loop factor times $M_s^2$,
according to eq.~(\ref{mu2R}).

\begin{figure}[htb]
\begin{center}
\includegraphics[width=7.5cm]{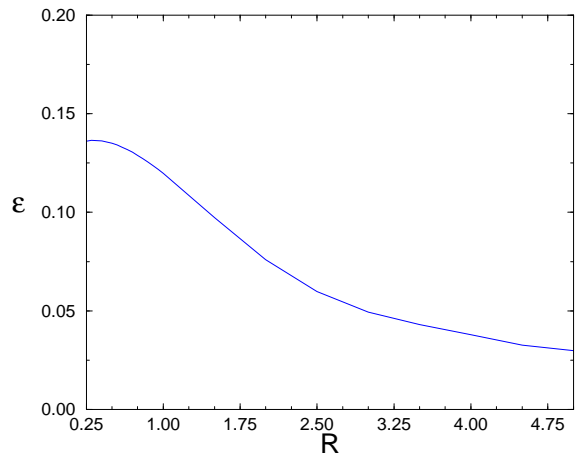}
\end{center}
\caption{\small The coefficient $\varepsilon$ of the one loop Higgs mass (\ref{mu2R}).
\label{epsilonfig}}
\end{figure}

More precisely, from the expression (\ref{strings}), one finds:
\be
\varepsilon^2(R) ={1\over 2\pi^2}\int_0^\infty \frac{dl}{\left(2\,
l\right)^{5/2}} 
{\theta_2^4\over 4\eta^{12}}\left(il+{1\over 2}\right) R^3
\sum_n n^2 e^{-2\pi n^2R^2l}\ ,
\label{epsilon2R}
\ee
which is plotted in Fig.~\ref{epsilonfig}.
For the asymptotic value $R\to 0$ (corresponding upon T-duality
to a large transverse dimension of radius $1/R$),
$\varepsilon(0)\simeq 0.14$, and the effective cut-off for the mass
term is $M_s$, as can be seen from Eq.~(\ref{mu2R}). At
large $R$, $\mu^2(R)$ falls off as $1/R^2$, 
which is the effective cut-off in the limit $R\to\infty$, as we argued above, 
in agreement with field theory
results in the presence of a compactified extra
dimension~\cite{SS, iadd}. In fact, in the
limit $R\to\infty$, an analytic approximation to $\varepsilon(R)$ gives:
\be
\varepsilon(R)\simeq \frac{\varepsilon_\infty}{M_s\, R}\, ,
\qquad\qquad
\varepsilon_\infty^2=\frac{3\, \zeta(5)}{4\, \pi^4}\simeq 0.008\, .
\label{largeR}
\ee

The potential (\ref{potencialh}) has the usual minimum, given by the
VEV of the neutral component of the Higgs doublet 
$v=\sqrt{-\mu^2/\lambda}$. Using the relation of $v$ with the $Z$
gauge boson mass, $M_Z^2=(g_2^2+g'^2)v^2/4$, and the expression
of the quartic coupling $\lambda$, 
one obtains for the Higgs mass a prediction which is the
MSSM value for $\tan\beta\to\infty$ and $m_A\to\infty$:
$m_H=M_Z$. 
The tree level Higgs mass is known to receive important
radiative corrections from the top-quark sector and rises
to values around 120 GeV. 
Furthermore, from (\ref{mu2R}), one can compute $M_s$ in terms of the Higgs
mass $m_H^2=-2\mu^2$:
\begin{equation}
M_s=\frac{m_H}{\sqrt{2}\, g\varepsilon}\, ,
\label{final}
\end{equation}
yielding naturally values in the TeV range.

\section{Standard Model on D-branes}
\vskip 0.2cm

The gauge group closest to the Standard Model one can easily obtain with 
D-branes is $U(3)\times U(2)\times U(1)$. The first factor arises from three
coincident ``color" D-branes. An open string with one end on
them is a triplet under $SU(3)$ and carries the same $U(1)$ charge for all three
components. Thus, the $U(1)$ factor of $U(3)$ has to be identified with {\it
gauged} baryon number. Similarly, $U(2)$ arises from two coincident ``weak"
D-branes and the corresponding abelian factor is identified with {\it gauged}
weak-doublet number. Finally, an extra $U(1)$ D-brane is necessary in 
order to accommodate the Standard Model without breaking the baryon number~\cite{akt}. 
In principle this $U(1)$ brane can be chosen to be
independent of the other two collections with its own gauge coupling. To improve
the predictability of the model, we choose to put it on top of either the
color or the weak D-branes~\cite{st}. In either case, the model has two independent gauge
couplings $g_3$ and $g_2$ corresponding, respectively, to the gauge groups $U(3)$
and $U(2)$. The $U(1)$ gauge coupling $g_1$ is equal to either $g_3$ or $g_2$.

Let us denote by $Q_3$, $Q_2$ and $Q_1$ the three $U(1)$ charges of $U(3)\times
U(2)\times U(1)$, in a self explanatory notation. Under $SU(3)\times SU(2)\times
U(1)_3\times U(1)_2\times U(1)_1$, the members of a family of quarks and
leptons have the following quantum numbers:
\ba
&Q &({\bf 3},{\bf 2};1,w,0)_{1/6}\nonumber\\
&u^c &({\bf\bar 3},{\bf 1};-1,0,x)_{-2/3}\nonumber\\
&d^c &({\bf\bar 3},{\bf 1};-1,0,y)_{1/3}\label{charges}\\
&L   &({\bf 1},{\bf 2};0,1,z)_{-1/2}\nonumber\\
&l^c &({\bf 1},{\bf 1};0,0,1)_1\nonumber
\ea
The values of the $U(1)$ charges $x,y,z,w$ will be fixed below so that
they lead to the right hypercharges, shown for completeness as subscripts.

\begin{figure}[htb]
\begin{center}
\includegraphics[width=8cm]{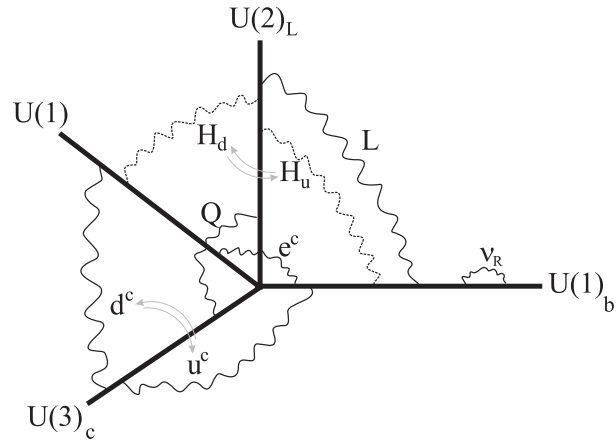}
\end{center}
\caption{\small A minimal Standard Model embedding on D-branes.
\label{SM}}
\end{figure}

It turns out that there are two possible ways of embedding the Standard Model
particle spectrum on these stacks of branes~\cite{akt}, which are shown pictorially in
Fig.~\ref{SM}.
The quark doublet $Q$ corresponds necessarily to a massless excitation of an
open string with its two ends on the two different collections of branes 
(color and weak). As seen from the figure, a fourth brane stack is needed
for a complete embedding, which is chosen to be a $U(1)_b$ extended in
the bulk. This is welcome since one can accommodate right handed neutrinos
as open string states on the bulk with sufficiently small Yukawa couplings
suppressed by the large volume of the bulk~\cite{Rnus}.
The two models are obtained by an exchange of the up and down 
antiquarks, $u^c$ and $d^c$, which correspond to open strings with one 
end on the color branes and the other either on the $U(1)$ brane, or on
the $U(1)_b$ in the bulk. The lepton doublet $L$ arises from an open string
stretched between the weak branes and $U(1)_b$, while the antilepton $l^c$
corresponds to a string with one end on the $U(1)$ brane and the other in the 
bulk. For completeness, we also show the two possible Higgs states $H_u$
and $H_d$ that are both necessary in order to give tree-level masses to
all quarks and leptons of the heaviest generation.

\begin{figure}[htb]
\begin{center}
\includegraphics[width=10cm]{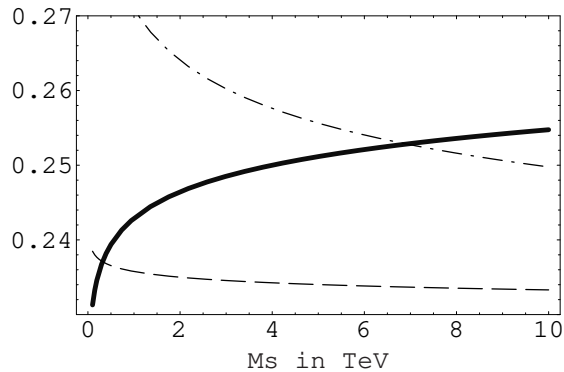}
\end{center}
\vspace{-1.0cm}
\caption{The experimental value of $\sin^2\theta_W$ (thick curve), and
the theoretical predictions (\ref{sintheta}).}
\label{sin}
\end{figure}

The weak hypercharge $Y$ is a linear combination of the three $U(1)$'s:
\be
Y= Q_1+{1\over 2}Q_2+c_3 Q_3\quad ;\quad c_3=-1/3\ {\rm or}\ 2/3\, ,
\label{Y}
\ee
where $Q_N$ denotes the $U(1)$ generator of $U(N)$ normalized so that
the fundamental representation of $SU(N)$ has unit charge. 
The corresponding $U(1)$ charges appearing in eq.~(\ref{charges}) are
$x=-1$ or $0$, $y=0$ or $1$, $z=-1$, and $w=1$ or $-1$, 
for $c_3=-1/3$ or $2/3$, respectively.
The hypercharge coupling $g_Y$ is given by~\footnote{The gauge couplings
$g_{2,3}$ are determined at the tree-level by the string coupling and other
moduli, like radii of longitudinal dimensions. In higher orders, they also
receive string threshold corrections.}:
\be
{1\over g_Y^2}={2\over g_1^2}+{4c_2^2\over g_2^2}+
{6c_3^2\over g_3^2}\, .
\label{gY}
\ee
It follows that the weak angle $\sin^2\theta_W$, is given by:
\be
\sin^2\theta_W\equiv{g_Y^2\over g_2^2+g_Y^2}=
{1\over 2+2g_2^2/g_1^2+6c_3^2g_2^2/g_3^2}\, ,
\label{sintheta}
\ee
where $g_N$ is the gauge coupling of $SU(N)$ and $g_1=g_2$ or $g_1=g_3$ 
at the string scale. In order to compare the theoretical predictions with the 
experimental value of $\sin^2\theta_W$ at $M_s$, we plot
in Fig.~\ref{sin} the corresponding curves as functions of $M_s$. 
The solid line is the experimental curve. The dashed line is the plot of the
function (\ref{sintheta}) for $g_1=g_2$ with $c_3=-1/3$ while the dotted-dashed
line corresponds to $g_1=g_3$ with $c_3=2/3$. The other two possibilities are
not shown because they lead to a value of $M_s$ which is too high to protect the
hierarchy. Thus, the second case, where the
$U(1)$ brane is on top of the color branes, is compatible with low energy data
for $M_s\sim 6-8$ TeV and $g_s\simeq 0.9$.

From Eq.~(\ref{sintheta}) and Fig.~\ref{sin}, we find the ratio of the
$SU(2)$ and $SU(3)$ gauge couplings at the string scale to be
$\alpha_2/\alpha_3\sim 0.4$. This ratio can be arranged by an appropriate
choice of the relevant moduli. For instance, one may choose the color and
$U(1)$ branes to be D3 branes while  the weak branes to be D7 branes.
Then, the ratio of couplings above can be explained by choosing the volume
of the four  compact dimensions of the seven branes to be $V_{4}=2.5$ in
string units. This being larger than one is consistent with the picture
above. Moreover it predicts an interesting spectrum of KK states for the
Standard model, different from  the naive choices that have appeared
hitherto: the only Standard Model particles that have KK descendants are
the W bosons as well as the hypercharge gauge boson. However, since the
hypercharge is a linear combination of the three $U(1)$'s, the massive $U(1)$
KK gauge bosons do not couple to the hypercharge but to the weak 
doublet number.

\section{Non-compact extra dimensions and localized gravity}
\vskip 0.2cm

There are several motivations to study localization of gravity in non-compact
extra dimensions: (i) it avoids the problem of fixing the moduli associated to 
the size of the compactification manifold; (ii) it provides a new approach to the
mass hierarchy problem; (iii) there are modifications of gravity at large distances
that may have interesting observational consequences.
Two types of models have been studied: warped metrics in curved space~\cite{RS}, 
and infinite size extra dimensions in flat space~\cite{DGP}.
The former, although largely inspired by stringy developments and having 
used many string-theoretic techniques, have not yet a clear and calculable 
string theory realization~\cite{VerlindeFY}. In any case,
since curved space is always difficult to handle in string theory, in the following 
we concentrate mainly on the latter, formulated in
flat space with gravity localized on a subspace of the bulk. It turns out that
these models of induced gravity have an interesting string theory 
realization~\cite{AMV} that we describe below, 
after presenting first a brief overview of the warped case~\cite{rev}.

\subsection{Warped spaces}
\vskip 0.2cm

In these models, space-time is a slice of anti de Sitter space (AdS) in $d=5$ 
dimensions while our universe forms a four-dimensional (4d) flat 
boundary~\cite{RS}. The corresponding line element is:
\be
ds^2=e^{-2k|y|}\eta_{\mu\nu}dx^\mu dx^\nu+dy^2\quad ;\quad
\Lambda=-24M^3k^2\, ,
\label{ds2}
\ee
where $M,\Lambda$ are the 5d Planck mass and cosmological constant,
respectively, and the parameter $k$ is the curvature of AdS$_5$.
The fifth coordinate $y$ is restricted on the interval $[0,\pi r_c]$.
Thus, this model
requires two `branes', a UV and an IR, located at the two end-points of the 
interval, $y=0$ and $y=\pi r_c$, respectively.
The vanishing of the 4d cosmological constant requires to fine tune
the two tensions: $T=-T^\prime=24M^3k^2$. The 4d Planck mass is
given by:
\be
M_P^2={1\over k}(1-e^{-2\pi kr_c})M^3\, .
\label{mp}
\ee

Note that the IR brane can move to infinity by taking the limit $r_c\to\infty$,
while $M_P$ is kept finite and thus 4d gravity is always present on the
brane. The reason is that the internal volume remains finite in the
non-compact limit along the positive $y$ axis. As a result, gravity is 
kept localized on the UV brane, while the Newtonian potential gets 
corrections, $1/r+1/k^2r^3$, which are identical with those arising in
the compact case of two flat extra dimensions. Using the experimental
limit $k^{-1}\simlt 0.1$ mm and the relation (\ref{mp}), one finds a 
bound for the 5d gravity scale $M\simgt 10^8$ GeV, corresponding
to a brane tension $T\simgt 1$ TeV. Notice that this bound is not valid
in the compact case of six extra dimensions, because their size is
in the fermi range and thus the $1/r^3$ deviations of Newton's law
are cutoff at shorter distances.

\subsection{The induced gravity model}
\vskip 0.2cm

The {\sc dgp}  model and its generalizations are specified by a bulk
Einstein-Hilbert ({\sc eh}) term and a four-dimensional {\sc eh} term~\cite{DGP}:
\begin{equation}
M^{2+n} \int_{{\cal M}_{4+n}} d^4x d^ny\, \sqrt{G}\, \mathcal{R}_{(4+n)} + 
M_P^2 \int_{{\cal M}_4} d^4x\sqrt{g}\, \mathcal{R}_{(4)}\, ;\quad
M_P^2\equiv{r_c^n M^{2+n}}
\label{e:dgp}
\end{equation}
with $M$ and $M_P$ the  (possibly independent) respective Planck
scales. The scale $M \geq 1$~TeV would be related to the short-distance scale
below which {\sc uv} quantum gravity or stringy effects are important. 
The four-dimensional metric is the restriction of the bulk metric
$g_{\mu \nu}=\left.G_{\mu \nu}\right|$ and we assume the {\sc
world}\footnote{We avoid calling ${\cal M}_4$ a brane because, as we will see
below, gravity localizes on singularities of the internal manifold, such as orbifold 
fixed points. Branes with localized matter can be introduced independently of
gravity localization.}
rigid, allowing the gauge $\left. G_{i\mu }\right|=0$ with $i\geq 5$. 
Finally, only intrinsic curvature
terms are omitted but no Gibbons--Hawking term is needed.

\subsubsection{Co-dimension one}

In the case of co-dimension one bulk ($n=1$) and $\delta$-function
localization, it is easy to see that $r_c$ is a crossover scale where gravity
changes behavior on the {\sc world}. Indeed, by 
Fourrier transform the quadratic part of the action (\ref{e:dgp}) with respect
to the 4d position $x$, at the {\sc world} position $y=0$, one obtains 
$M^{2+n}(p^{2-n}+r_c^n p^2)$, where $p$ is the 4d momentum. It follows
that for distances smaller than $r_c$ (large momenta), the first term becomes
irrelevant and the graviton propagator on the ``brane" exhibits
four-dimensional behavior $(1/p^2)$ with Planck constant $M_P=M^3r_c$.
On the contrary, at large distances, the first term becomes dominant and the
graviton propagator acquires a five-dimensional fall-off $(1/p)$ 
with Planck constant $M$. Imposing $r_c$ to be larger than the size 
of the universe, $r_c\simgt 10^{28}$ cm, one finds $M\simlt 100$ MeV, 
which seems to be in conflict with experimental bounds. However, there
were arguments that these bounds can be evaded, even for values of
the fundamental scale $M^{-1}\sim 1$ mm that one may need for suppressing 
the quantum  corrections of the cosmological constant~\cite{DGP}.

On the other hand, in the presence of non-zero brane
thickness $w$, a new crossover length-scale seems to appear,
$R_c\sim(wr_c)^{1/2}$~\cite{KTT} or $r_c^{3/5}w^{2/5}$~\cite{Luty:2003vm}.
\begin{center}
\vskip 0.5cm
\begin{tabular}{ccc|ccccccc|cccccc|ccc}
&~ ~ & ~ ~ & ~ ~ & ~ ~ & ~ ~ & ~ ~ & ~ ~ & ~ ~ & ~ ~ & ~ ~ & ~ ~ & ~ ~ & {4d} 
& ~ ~  & ~ ~ & ~ ~ & {5d} & ~ ~ \\
\hline
& & & & & & & & & & & & & & &  & & & \\
\end{tabular}
\end{center}
$~$\hskip 2.5cm $w$\hskip 2.9cm $R_c$\hskip 2.4cm $r_c$\vskip -0cm
$~$\hskip 1.6cm {$\nearrow$\hskip 1.8cm $\uparrow$}\vskip -0cm
$~${\small UV cutoff\hskip 1.8cm 5d or strong coupling}\vskip 0.5cm
\noindent
Below this scale, the theory acquires either again a five-dimensional behavior,
or a strong coupling regime. For $r_c\sim 10^{28}$ cm, the new crossover
scale is of order $R_c\sim 10^{-4}-10$ m.

\subsubsection{Higher co-dimension}

The situation changes drastically for more than one non-compact bulk
dimensions, $n>1$, due to the ultraviolet properties of the higher-dimensional
theories. Indeed, from the action (\ref{e:dgp}),
the effective potential between two test masses in four dimensions
\begin{eqnarray}
\label{e:Force}
\int [d^3x] \, e^{-i p\cdot x} \, V(x)&=&\frac{D(p)}{1+ r_c^n\, p^2 \, D(p)}  \, \left[\tilde T_{\mu \nu}
T^{\mu \nu}- \frac{1}{2+n} \, \tilde T^\mu _\mu \, T^\nu_\nu\right]\\
D(p) &=& \int [d^nq] \, \frac{f_w(q)}{p^2+q^2}
\label{e:Dp}
\end{eqnarray}
is a function of the bulk graviton retarded Green's function
$G(x,0;0,0)=\int [d^4p] \, e^{i p\cdot x}\, D(p)$ evaluated for two
points localized on the {\sc world} ($y=y'=0$).  The
integral~(\ref{e:Dp}) is {\sc uv}-divergent for $n>1$ unless there is
a non-trivial brane thickness profile $f_w(q)$ of width $w$. If the
four-dimensional {\sc world} has zero thickness, $f_w(q)\sim 1$, the bulk
graviton does not have a normalizable wave function. It therefore cannot contribute
to the induced potential, which always takes the form  $V(p) \sim 1/p^2$ and
Newton's law remains four-dimensional at all distances. 

For a non-zero thickness $w$, there is only one crossover length scale, $R_c$: 
\begin{equation}
\label{e:Rc}
R_c=w\left(\frac{r_c}{w}\right)^{\frac{n}{2}}\, , 
\end{equation}
above which one obtains a higher-dimensional
behaviour~\cite{Dvali:2000xg}.
Therefore the effective
potential presents two regimes: (i) at short distances ($w\ll r\ll R_c$) the
gravitational interactions are mediated by the localized four-dimensional
graviton and Newton's potential on the {\sc world} is given by $V(r)\sim
1/r$ and, (ii) at large distances ($r\gg R_c$) the modes of the bulk graviton
dominate, changing the potential. Note that for $n=1$ the expressions~(\ref{e:Force})
and~(\ref{e:Dp}) are finite and unambiguously give $V(r)\sim 1/r$ for
$r\gg r_c$. For a co-dimension bigger than 1, the precise behavior for
large-distance interactions depends \emph{crucially} on the {\sc uv}
completion of the theory.
\begin{center}
\vskip 0.5cm
\begin{tabular}{cccccccc|cccccccc}
&~ ~ &~ ~ & {4d} &~ ~ &~ ~  &~ ~ & ~ ~ & ~ ~ & ~ ~ &  {higher d} & ~ ~ & ~ ~ & \\
\hline
& & & & & & & & & & & & & \\
\end{tabular}
\end{center}
$~$\hskip 5.7cm $R_c$\vskip 0.5cm
\noindent
At this point we stress a fundamental difference with the \emph{finite extra
dimensions} scenarios. In these cases Newton's law gets higher-dimensional
at distances smaller than the characteristic size of the extra dimensions. This
is precisely the opposite of the case of infinite volume extra dimensions that
we discuss here.

As mentioned above, for higher co-dimension, there is an interplay between
UV regularization and IR behavior of the theory. Indeed, several works
in the literature raised unitarity~\cite{DR} and strong coupling
problems~\cite{Rubakov:2003zb} which 
depend crucially on the {\sc uv} completion of the theory.
A unitary {\sc uv} regularization for the higher co-dimension version
of the model has been proposed in~\cite{Kolanovic:2003am}. 
It would be interesting to address these questions in a precise string theory context.
Actually, using for UV cutoff on the ``brane" the 4d Planck length $w\sim l_P$,
one gets for the crossover scale (\ref{e:Rc}): $R_c\sim M^{-1}(M_P/M)^{n/2}$.
Putting $M\simgt 1$ TeV leads to $R_c\simlt 10^{8(n-2)}$ cm.
Imposing $R_c\simgt 10^{28}$ cm, one then finds that the number of extra 
dimensions must be at least six, $n\ge 6$, which is realized nicely in string theory
and provides an additional motivation for studying possible string theory realizations.

\subsection{String theory realization}
\vskip 0.2cm

In the following, we explain how to realize the gravity induced model
(\ref{e:dgp}) with $n\geq6$
as the low-energy effective action of string theory on a non-compact
six-dimensional manifold $\mathcal{M}_6$~\cite{AMV}. We work in the context of
$\mathcal{N}=2$ supergravities in four dimensions but the mechanism for
localizing gravity is independent of the number of supersymmetries. Of course
for  $\mathcal{N}\geq 3$ supersymmetries, there is no localization. We also
start with the compact case and take the decompactification limit. The localized
properties are then encoded in the different volume dependences.

In string perturbation, corrections to the four-dimensional Planck mass are
in general very restrictive. In the heterotic string, they vanish to
all orders in perturbation theory~\cite{AGN}; in type {\sc I} theory,
there are moduli-dependent corrections generated by open
strings~\cite{AB}, but they vanish when the
manifold $\mathcal{M}_6$ is decompactified; in type {\sc II} theories, they are constant,
independent of the moduli of the manifold $\mathcal{M}_6$, and receive
contributions only from tree and one-loop levels that we describe below 
(at least for supersymmetric backgrounds)~\cite{AMV,AFMN}.
Finally, in the context of M-theory, one obtains a similar localized action of 
gravity kinetic terms in five dimensions, corresponding to the strong coupling limit
of type IIA string~\cite{AMV}.

The origin of the two {\sc eh} terms in~(\ref{e:dgp}) can be traced back to the 
perturbative corrections to the eight-derivative effective
action of type~{\sc II} strings in ten dimensions. These corrections include 
the tree-level and one-loop terms given 
by:\footnote{The rank-eight tensor $t_8$ is defined as
$t_8M^4\equiv -6({\rm tr}M^2)^2 +24{\rm tr}M^4$. 
See~\cite{Peeters:2000qj} for more details.}
\begin{eqnarray}
\label{e:Action}
\frac{1}{l_s^8}& &\int_{M_{10}} \frac{1}{g_s^2} \mathcal{R}_{(10)} +
\frac{1}{l_s^2}\int_{M_{10}}\,  \left(
\frac{2\zeta(3)}{g_s^2} +  4\zeta(2)\right)t_8t_8 R^4 \\
&-&\frac{1}{l_s^2} \int_{M_{10}} \left( \frac{2\zeta(3)}{g_s^2}
\mp 4\zeta(2)\right) R\wedge R \wedge R\wedge R \wedge e \wedge e
+\cdots\nonumber{}
\end{eqnarray}
where $\phi$ is the dilaton field determining the string coupling 
$g_s=e^{\langle\phi\rangle}$, and the $\pm $ sign corresponds to the 
type~{\sc iia/b} theory.

On a direct product space-time $\mathcal{M}_6 \times \mathbb{R}^4$, the 
$t_8t_8 R^4$ contribute in four dimensions to $R^2$ and $R^4$ terms~\cite{AFMN}.
At the level of zero modes, the second $R^4$ term in (\ref{e:Action}) splits as: 
\be
\int_{M_6} \, R\wedge R \wedge R \times 
\int_{M_4} \, \mathcal{R}_{(4)}=\chi\, \int_{M_4}\, \mathcal{R}_{(4)}\, ,
\ee
where $\chi$ is the Euler number of the $M_6$ compactification manifold.
We thus obtain the action terms:
\begin{equation}
\label{e:Local}
\frac{1}{l_s^8}\int_{M_4 \times   M_6}\, \frac{1}{g_s^2}\, \mathcal{R}_{(10)} + 
\frac{\chi}{l_s^2} \int_{M_4}\, \left(-\frac{2\zeta(3)}{g_s^2} 
\pm   4\zeta(2)\right) \mathcal{R}_{(4)}\, ,
\end{equation}
which gives the expressions for the Planck masses $M$ and $M_p$:
\be
M^2\sim M_s^2/g_s^{1/2}\quad;\quad 
M_P^2\sim\chi ({c_0\over g_s^2}+c_1)M_s^2\, ,
\label{relations}
\ee
with $c_0=-2\zeta(3)$ and $c_1=\pm 4\zeta(2)=\pm 2\pi^2/3$.

It is interesting that the appearance of the induced 4d localized term preserves
${\cal N}=2$ supersymmetry and is independent of the localization mechanism
of matter fields (for instance on D-branes). Localization requires the internal
space $M_6$ to have a non-zero Euler characteristic $\chi\neq 0$. Actually,
in type {\sc iia/b} compactified on a Calabi-Yau manifold, 
$\chi$ counts the difference between the numbers of
$\mathcal{N}=2$ vector multiplets and hypermultiplets: $\chi=\pm 4(n_V-n_H)$
(where the graviton multiplet counts as one vector). Moreover, in the 
non-compact limit, the Euler number can in general split in different singular 
points of the internal space, $\chi=\sum_I\chi_I$, giving rise
to different localized terms at various points $y_I$ of the internal space.
A number of conclusions (confirmed by string calculations in~\cite{AMV}) 
can be reached by looking closely at~(\ref{e:Local}):

$\triangleright$  $M_p\gg M$ requires a large non-zero Euler
characteristic for $M_6$, and/or a weak string coupling constant $g_s\to0$.

$\triangleright$ Since $\chi$ is a topological invariant the localized
$\mathcal{R}_{(4)}$ term coming from the closed string sector is universal,
independent of the background geometry and dependent only on the internal 
topology\footnote{Field theory
computations of~\cite{Adler:1982ri} show that the Planck mass
renormalization depends on the {\sc uv} behavior of the matter fields
coupling to the external metric. But, even in the
supersymmetric case, the corrections are not obviously given by an index.}.  
It is a matter of simple inspection to see that if one wants to have a localized 
{\sc eh} term in less than ten dimensions, namely something linear in curvature, 
with non-compact internal space in all directions, \emph{the only possible
dimension is four} (or five in the strong coupling M-theory limit).

$\triangleright$ In order to find the width $w$ of the localized term, 
one has to do a separate analysis. On general grounds, using 
dimensional analysis in the limit $M_P\to\infty$, one expects the effective
width to vanish as a power of $l_P\equiv M_P^{-1}$: 
$w\sim l_P^\nu/l_s^{\nu-1}$ with $\nu>0$.
The computation of $\nu$ for a general Calabi-Yau space, besides
its technical difficulty, presents an additional important complication:
from the expression (\ref{relations}), $l_P\sim g_sl_s$ in the weak 
coupling limit. Thus, $w$ vanishes in perturbation theory and one has 
to perform a non-perturbative analysis to extract its behavior. Alternatively, 
one can examine the case of orbifolds. In this limit, $c_0=0$, $l_P\sim l_s$,
and the hierarchy $M_P>M$ is achieved only in the limit of large $\chi$.

The one-loop graviton amplitude for the supersymmetric
orbifold $T^6/\mathbb{Z}_N$, takes the form of a sum of quasi-localized
contributions at the positions of the fixed points $x_f$ of the
orbifold~\cite{AMV}:
\begin{equation}
\langle V_g^3 \rangle \sim \frac{1}{N} \sum_{(h,g)}\sum_{x_f} \,  \int_{\mathcal{F}}
\frac{d^2\tau}{\tau_2^2}\, \int \prod_{i=1}^3 \frac{d^2z_i}{\tau_2}\,
\frac{1}{F_{(h,g)}(\tau,z_i)^3}\, e^{- \frac{(y-x_f)^2}{\alpha'F_{(h,g)}(\tau,z_i)}}\, ,
\label{3point}
\end{equation}
where $(h,g)$ denote the orbifold twists and $\tau=\tau_1+i\tau_2$ is the
complex modulus of the world-sheet torus, integrated over its fundamental
domain $\cal F$. The above expression (\ref{3point}) gives the three-point
amplitude involving three 4d gravitons on-shell.
Focusing on one particular fixed point $x_f=0$ and sending the radii to
infinity, we obtain the effective action for the quasi-localized {\sc eh} term
\begin{equation}
\chi \, \int d^4x d^6y \sqrt{g} f_w(y) \, \mathcal{R}_{(4)}\, 
\end{equation}
with a width given by the four-dimensional induced Planck mass
\begin{equation}
\label{e:width}
w\simeq l_P  = l_s\, \chi^{-1/2}\, ,
\end{equation}
and the power $\nu=1$.

\subsubsection{Summary of the results}

Using $w\sim l_P$ and the relations (\ref{relations}) in the weak coupling limit
(with $c_0\ne 0$),
the crossover radius of eq.~(\ref{e:Rc}) is given by the string parameters ($n=6$)
\begin{equation}
\label{e:RcS}
R_c = \frac{r_c^3}{w^2}\sim g_s\frac{l_s^4}{l_P^3} \simeq g_s\times 10^{32}\ {\rm cm}\, ,
\end{equation}
for $M_s\simeq 1$ TeV.  Because $R_c$ has to be of cosmological size, the
string coupling can be relatively small, and the Euler number
$|\chi|\simeq g_s^2\, l_P \sim g_s^2\times 10^{32}$ must be very large.  
The hierarchy is obtained mainly thanks to the large value of $\chi$, 
so that lowering the bound on $R_c$ lowers the value of $\chi$.  
Our actual knowledge of gravity at very large distances
indicates~\cite{Lue} that $R_c$ should be of the order of the Hubble radius 
$R_c \simeq 10^{28}$ cm, which implies $g_s \geq 10^{-4}$ and $|\chi|\simgt 10^{24}$.  
A large Euler number implies only a large number of closed string
massless particles with no a-priori constraint on the observable gauge and
matter sectors, which can be introduced for instance on D3-branes placed at the
position where gravity localization occurs. All these particles are localized
at the orbifold fixed points (or where the Euler number is concentrated in the
general case), and should have sufficiently suppressed
gravitational-type couplings, so that their presence with such a huge
multiplicity does not contradict observations.  Note that these results depend
crucially on the scaling of the width $w$ in terms of the Planck length:
$w\sim l_P^\nu$, implying $R_c\sim 1/l_P^{2\nu +1}$ in string units. If there
are models with $\nu>1$, the required value of $\chi$ will be much lower,
becoming $\mathcal{O}(1)$ for $\nu\ge 3/2$. In this case, the hierarchy could
be determined by tuning the string coupling to infinitesimal values, 
$g_s\sim 10^{-16}$.

The explicit string realization of localized induced gravity models offers a consistent
framework that allows to address a certain number of interesting physics
problems. In particular, the effective UV cutoff and the study of the gravity
force among matter sources localized on D-branes. It would be also interesting to 
perform explicit model building and study in detail the
phenomenological consequences of these models and compare to other
realizations of TeV strings with compact dimensions.

\section*{Acknowledgments}
\vskip 0.2cm

This work was supported in part by the European Commission under the
RTN contract MRTN-CT-2004-503369, and in part by the INTAS
contract 03-51-6346.


\printindex
\end{document}